\documentclass[runningheads,a4paper]{llncs}

\newif\ifselfarchive
\selfarchivefalse

\title{Validating Traces of Distributed Programs\\Against \tlaplus Specifications%
  \thanks{%
    This work was partly supported by a grant from Oracle Corporation.
    \ifselfarchive\\
    This version of the publication has been accepted for publication,
    after peer review but is not the Version of Record. The Version of Record is
    available online at \protect{\url{http://dx.doi.org/000000}}.
    Use of this Accepted Version is subject to the publisher's Accepted
    Manuscript terms of use
    \protect{\url{https://www.springernature.com/gp/open-research/policies/accepted-manuscript-terms}}.
    \else\relax\fi}}
\titlerunning{Trace validation for \tlaplus}
\author{%
  Horatiu Cirstea\inst{1} \and
  Markus A. Kuppe\inst{2} \and
  Benjamin Loillier\inst{1} \and
  Stephan Merz\inst{1}
}
\authorrunning{H. Cirstea, M. Kuppe, B. Loillier, S. Merz}
\institute{
  University of Lorraine, CNRS, Inria, LORIA, Nancy, France
  \and
  Microsoft Research
}

\usepackage{capt-of}
\usepackage{placeins}
\usepackage{multirow}
\usepackage{xspace}
\usepackage{latexsym}
\usepackage{amssymb}
\usepackage{xcolor}
\usepackage[T1]{fontenc}
\usepackage{relsize}
\usepackage{enumitem}
\usepackage{pifont}
\usepackage{xcolor,soul,framed} 
\usepackage[cmex10]{amsmath}
\usepackage{array}
\usepackage{mdwmath}
\usepackage{mdwtab}
\usepackage{eqparbox}
\usepackage{url}
\usepackage{pcallst}
\usepackage{javalst}
\usepackage{jsonlst}
\usepackage{tikz}
\usepackage{subfig}
\usetikzlibrary{automata,calc,shapes,positioning}

\definecolor{dkgreen}{rgb}{0,0.6,0}
\definecolor{orange}{rgb}{1.0,0.49,0.0}
\definecolor{bluegray}{rgb}{0.4, 0.6, 0.8}
\definecolor{dpcolor}{rgb}{0.58,0,0.82}
\definecolor{javacolor}{rgb}{0,0,0} 
\definecolor{pcalcolor}{rgb}{0,0,0}  

\newcommand{\para}[1]{\smallskip\noindent\emph{#1}}
\newcommand{\tlaplus}{TLA\textsuperscript{\textup{+}}\xspace}

\newcommand{\deq}{\mathrel{\smash{\stackrel{\scriptscriptstyle\Delta}{=}}}}
\newcommand{\seq}[1]{\langle #1 \rangle}
\newcommand{\str}[1]{{\textcolor{blue}{\texttt{"#1"}}}}

\newcommand{\ie}[0]{i.e.}

\newcommand{\TM}{{TM}}
\newcommand{\RMs}{{RMs}}
\newcommand{\RM}{{RM}}

\newcommand{\bTP}[1]{{TP, $#1$ RMs}}
\newcommand{\bKVS}[3]{{KV, #1a, #2k, #3v}} 

\newcommand{\JSON}{\textsf{JSON}}
\newcommand{\Java}{\textsf{Java}}

\newcommand{\eg}{e.g.}

\definecolor{hc}{rgb}{0.58,0,0.82}
\definecolor{sm}{rgb}{0,0.5,0}

\makeatletter

\newcounter{abr@ctr}
\newcommand{\abr@c}{\c@abr@ctr\advance\c@abr@ctr\@ne}

\ifx\documentclass\undefined
\else
  \DeclareSymbolFont{tlaitalics}{\encodingdefault}{cmr}{m}{it}
  \let\itfam\symtlaitalics
\fi

\newcommand{\noTeXmath}{%
\c@abr@ctr=\itfam
\multiply\c@abr@ctr"100\relax
\advance\c@abr@ctr "7061\relax
\mathcode`a=\abr@c\mathcode`b=\abr@c\mathcode`c=\abr@c\mathcode`d=\abr@c
\mathcode`e=\abr@c\mathcode`f=\abr@c\mathcode`g=\abr@c\mathcode`h=\abr@c
\mathcode`i=\abr@c\mathcode`j=\abr@c\mathcode`k=\abr@c\mathcode`l=\abr@c
\mathcode`m=\abr@c\mathcode`n=\abr@c\mathcode`o=\abr@c\mathcode`p=\abr@c
\mathcode`q=\abr@c\mathcode`r=\abr@c\mathcode`s=\abr@c\mathcode`t=\abr@c
\mathcode`u=\abr@c\mathcode`v=\abr@c\mathcode`w=\abr@c\mathcode`x=\abr@c
\mathcode`y=\abr@c\mathcode`z=\abr@c
\c@abr@ctr=\itfam
\multiply\c@abr@ctr"100\relax
\advance\c@abr@ctr "7041\relax
\mathcode`A=\abr@c\mathcode`B=\abr@c\mathcode`C=\abr@c\mathcode`D=\abr@c
\mathcode`E=\abr@c\mathcode`F=\abr@c\mathcode`G=\abr@c\mathcode`H=\abr@c
\mathcode`I=\abr@c\mathcode`J=\abr@c\mathcode`K=\abr@c\mathcode`L=\abr@c
\mathcode`M=\abr@c\mathcode`N=\abr@c\mathcode`O=\abr@c\mathcode`P=\abr@c
\mathcode`Q=\abr@c\mathcode`R=\abr@c\mathcode`S=\abr@c\mathcode`T=\abr@c
\mathcode`U=\abr@c\mathcode`V=\abr@c\mathcode`W=\abr@c\mathcode`X=\abr@c
\mathcode`Y=\abr@c\mathcode`Z=\abr@c}

\makeatother
\noTeXmath

\begin{document}
\date{}
\maketitle

\begin{abstract}
  \tlaplus is a formal language for specifying systems, including 
  distributed algorithms, that is supported by powerful verification tools.
  In this work we present a framework for relating traces of distributed programs 
  to high-level specifications written in \tlaplus. The problem is reduced to 
  a constrained model checking problem, realized using the TLC model checker.
  Our framework consists of an API for instrumenting Java programs in order to 
  record traces of executions, of a collection of \tlaplus operators that are 
  used for relating those traces to specifications, and of scripts for running 
  the model checker. Crucially, traces only contain updates to 
  specification variables rather than full values, and developers may choose
  to trace only certain variables. 
  We have applied our approach to several distributed
  programs, detecting 
  discrepancies between the specifications and the implementations in all cases. 
  We discuss reasons for these discrepancies, 
  best practices for instrumenting programs,
  and how to interpret the verdict produced by TLC.
\end{abstract}

\section{Introduction}
\label{sec:intro}

Distributed systems are at the heart of modern cloud services and they 
are known to be error-prone, due to phenomena such as message delays or failures 
of nodes and communication networks. Applying formal methods during the design 
and development of these systems can help increase the confidence in their 
correctness and resilience. For example, the \tlaplus \cite{lamport:specifying}
specification language and verification tools have been successfully used 
in industry \cite{Newcombe2015,schultz:logless} for designing distributed 
algorithms underlying modern cloud systems. \tlaplus and similar specification 
formalisms are most useful for describing and analyzing systems at high 
levels of abstraction, but they do not provide much support for validating actual 
implementations of these systems. Although \tlaplus supports a 
notion of refinement, formally proving a chain of refinements from a 
high-level design of a distributed algorithm to an actual implementation 
would be a daunting task, complicated by the fact that standard programming 
languages do not provide explicit control of the grain of atomicity of the 
running program. In this work, we present a lightweight approach to validating 
distributed programs against high-level specifications that relies on recording 
finite traces of program executions and leveraging the \tlaplus model checker TLC 
\cite{yu:tlc} for comparing those traces to the state machine described in 
the \tlaplus specification. Although this approach does not provide formal 
correctness guarantees, even when the \tlaplus specification has been 
extensively verified, we have found it very useful for discovering and 
analyzing discrepancies between the runs of distributed programs and their 
high-level specifications. We have thus been able to discover 
serious bugs that had gone undetected by more traditional quality assurance techniques.

\begin{figure}[tb]
  \centering
  \includegraphics[width=\textwidth]{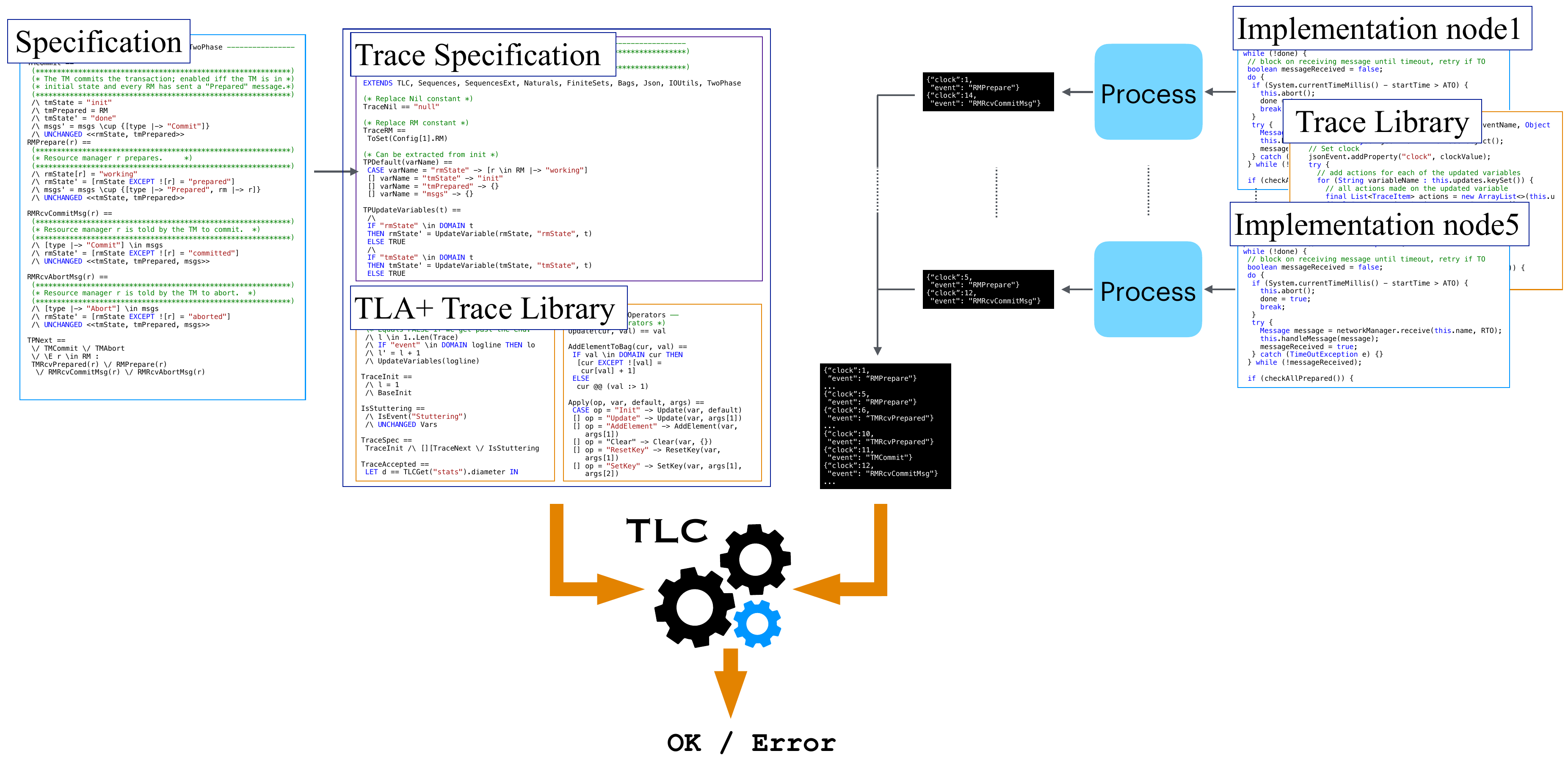}
  \caption{Overview of trace validation.}
  \label{fig:overview}
\end{figure}

Figure~\ref{fig:overview} summarizes the approach. The starting point
consists of a distributed \textsf{Implementation} and of the
\textsf{Specification}, written in \tlaplus, it is supposed to implement. 
We have designed a \textsf{Trace Library} that facilitates the instrumentation of
{\Java} programs in order to record information on how program
operations correspond to transitions described in the \tlaplus specification,
including updates of its variables. Executing the
instrumented programs produces traces in {\JSON} format that are
aggregated into a single file. We also provide a {\tlaplus} \textsf{Trace
  Library} 
that helps write a \textsf{Trace
  Specification}, extending the original specification. The model
checker TLC is then used to check the program trace against the trace specification.
Although our trace library was developed for {\Java}, the overall approach can
easily be adapted to any other imperative language. We also implemented a few 
features in TLC that support our approach.

Because the traces record the evolution of the state of the \tlaplus
specification, our approach is easiest to apply when the specification
exists prior to building the implementation, the implementor is
familar with it, and uses it as a blueprint when writing the
code. However, we have also used the approach in order to ``reverse
engineer'' a \tlaplus specification from an existing distributed
program and better understand its operation. Trace validation can also
help ensure that the specification and implementation remain in sync
over time because it is easy to apply it again in case of changes to the specification
or the implementation.

The main problem when instrumenting a program is to identify suitable 
``linearization points'' at which the program completes a step that 
corresponds to an atomic transition of the high-level state machine.
Basic guiding principles are to log an event when shared state has been 
updated, such as when sending or receiving messages, performing operations 
on locks or on stable data storage. Feedback from trace validation can 
help with adapting the instrumentation in order to take into account 
different grain of atomicity between the specification and the implementation,
as discussed later.
Because data representation generally
differs between the \tlaplus specification and the actual program, it may be 
difficult or impractical to compute the value of a specification variable 
(or its update) corresponding to the data manipulated by the implementation.
We therefore allow traces to be incomplete and only record some information 
about the corresponding abstract state. We reduce the problem of trace 
validation to one of constrained model checking and show how TLC can 
reconstruct missing information. This leads to a tradeoff between the 
precision of information recorded in the trace (and potentially of the
verdict of validation) and the amount of search that TLC must perform during 
model checking.

The paper\footnote{This is an extended version of~\cite{traceValidationSEFM2024}.} is organized as follows: Section~\ref{sec:background} provides 
some background on \tlaplus and introduces our running example.
Our approach to instrumentation is 
described in Section~\ref{sec:instrument}. In Section~\ref{sec:verify}
we formalize the trace validation problem, describe how we realized the 
approach using TLC, and discuss our experience with it. 
Section~\ref{sec:related} discusses related work, and 
Section~\ref{sec:conclusion} concludes the paper and presents
some perspectives for future work.

\section{Background}
\label{sec:background}

\subsection{\tlaplus Specifications}
\label{sec:tlaplus}

\tlaplus~\cite{lamport:specifying} is a specification language based on
Zermelo-Fraenkel set theory and linear-time temporal logic that has found 
wide use for writing high-level specifications of concurrent and distributed
algorithms. It emphasizes the use of mathematical descriptions based on sets and
functions for specifying data structures. In \tlaplus,
the state space of a system is represented using variables,
and formulas are evaluated over \emph{behaviors}, i.e., sequences of 
states that assign values to variables. Algorithms are described as state
machines whose specifications are written in the canonical form
\(
  \mathit{Init} \land \Box[\mathit{Next}]_{\mathit{vars}} \land \mathit{L}.
\)
In this formula, $\mathit{Init}$ is a state predicate describing the possible initial
states of the system, $\mathit{Next}$ represents the next-state relation, 
usually written as the
disjunction of actions describing the possible state transitions, $\mathit{vars}$ is a
tuple containing all state variables that appear in the specification, and $\mathit{L}$
is a temporal formula asserting liveness and fairness assumptions. A state
predicate is a formula of first-order logic that
is evaluated over single states. A transition predicate (or, synonymously,
action) is a first-order formula that may contain unprimed and primed
occurrences of variables. Such a formula is evaluated over a pair of
states, with unprimed variables referring to the values before the transition
and primed variables to the values after the transition. The formula
$[\mathit{Next}]_{\mathit{vars}}$ holds of a pair of states $\seq{s,t}$ 
if either $\mathit{Next}$ holds
of $\seq{s,t}$ (and therefore the pair represents an actual step of the system)
or the tuple $\mathit{vars}$ evaluates to the same value in the two states (and
the pair represents a stuttering step). Systematically allowing for stuttering
steps enables the refinement of a specification~$S$ by another
specification~$I$ written at a lower level of abstraction to be represented as the validity of the implication $I \Rightarrow S$.
The complementary property $L$ is used
to express fairness assumptions and is at the basis of verifying liveness
properties of algorithms. Since in this work we only analyze finite traces of
programs, we ignore liveness properties and are interested in finite behaviors,
i.e., sequences $s_0 \dots s_n$ of states such that $\mathit{Init}$ holds of $s_0$
and $[\mathit{Next}]_{\mathit{vars}}$ holds for all pairs 
$\seq{s_i, s_{i+1}}$ for $i \in 0\,..\,n-1$.

\begin{figure}[!t]
\begin{pluscal}
CONSTANT RM
VARIABLES rmState, tmState, tmPrepared, msgs
vars == <<rmState, tmState, tmPrepared, msgs>>
Messages == [type : {"Prepared"}, rm : RM] \union [type : {"Commit", "Abort"}]
TPInit == 
   /\ rmState = [r \in RM |-> "working"] /\ tmState = "init" 
   /\ tmPrepared = {} /\ msgs = {}
RMPrepare(r) == 
   /\ UNCHANGED <<tmState, tmPrepared>> /\ rmState[r] = "working" 
   /\ rmState' = [rmState EXCEPT ![r] = "prepared"]
   /\ msgs' = msgs \union {[type |-> "Prepared", rm |-> r]}
RMRcvCommitMsg(r) == 
   /\ UNCHANGED <<tmState, tmPrepared, msgs>> /\ [type |-> "Commit"] \in msgs 
   /\ rmState' = [rmState EXCEPT ![r] = "committed"]
RMRcvAbortMsg == ...
TMRcvPrepared(r) == 
   /\ UNCHANGED <<rmState,tmState,msgs>> /\ tmPrepared' = tmPrepared \union {r}
   /\ tmState = "init" /\ [type |-> "Prepared", rm |-> r] \in msgs
TMCommit == 
   /\ UNCHANGED <<rmState, tmPrepared>> /\ tmState = "init" /\ tmPrepared = RM
   /\ tmState' = "done" /\ msgs' = msgs \union {[type |-> "Commit"]}
TMAbort == ...
TPNext ==  
   \/ TMCommit \/ TMAbort
   \/ \E r \in RM : \/ RMPrepare(r) \/ TMRcvPrepared(r) 
                 \/ RMRcvCommitMsg(r) \/ RMRcvAbortMsg(r)
Spec == TPInit /\ [][TPNext]_vars     
Consistent ==
   \A r1,r2 \in RM: ~(rmState[r1] = "aborted" /\ rmState[r2] = "committed")
\end{pluscal}
   \caption{\tlaplus Specification of Two-Phase Commit.}
   \label{fig:tp-tla}
\end{figure}

As a running example for this paper, Fig.~\ref{fig:tp-tla} contains an
excerpt of the \tlaplus specification of the well-known Two-Phase
Commit protocol where a transaction manager (\TM) helps a set of
resource managers (\RMs) reach agreement on whether to commit or abort
a transaction: the {\RM}s send messages indicating that
they
are prepared to commit, while the {\TM} listens for such messages and  
based on the votes of the participants
broadcasts a commit or an abort message. This
specification is part of a collection of example \tlaplus
modules.\footnote{\url{https://github.com/tlaplus/Examples/tree/master/specifications/transaction_commit};
the complete specification is also given in the appendix.}
The module first declares a constant parameter \inlpluscal{RM} that represents
the set of RMs and four variables representing the control states of the RMs
(represented as a function with domain \inlpluscal{RM}) and of the TM, the set
of RMs that have declared their preparedness to carry out the transaction, and
the set of messages that have been sent during the protocol. The initial state
of the system is described by the predicate \inlpluscal{TPInit}: every RM is in
state \str{working}, the TM in state \str{init}, and the sets of prepared RMs
and of messages are empty.
The following operators define actions that describe individual state
transitions. For example, \inlpluscal{RMPrepare(r)} represents an
RM~\inlpluscal{r} declaring its preparedness to carry out the
transaction by moving to control state \inlpluscal{"prepared"} and adding 
a corresponding message to the set of messages \inlpluscal{msgs}.
The action \inlpluscal{TPNext} corresponds to the next-state relation of
the state machine, defined as the disjunction of the previously defined actions,
and formula \inlpluscal{Spec} represents the overall specification.
The \tlaplus tools, including the model checker TLC and the proof assistant TLAPS \cite{Cousineau2012}, 
can be used to verify properties of the 
specification, including the invariant \inlpluscal{Consistent} that \RMs{} must agree about 
committing or aborting a transaction.

\subsection{{\Java} Implementation}
\label{sec:implementation}

A possible {\Java} implementation of the resource managers is
presented in Fig.~\ref{fig:tp-java}. Only a simplified version of
the main method is shown, the auxiliary methods are faithful
{\Java} translations of the actions in the {\tlaplus}
specification.\footnote{The full implementation
is available at
\url{https://github.com/lbinria/TwoPhase}.
The main method of the {\TM} is given in the appendix.}

\begin{figure}[t]
   \begin{javap}
public class ResourceManager {  
  String name, tmName;
  ResourceManagerState state;
  NetworkManager network;
  public void run() throws IOException {
     working();
     while (true) {
        sendPrepared();
        try {
           Message message = network.receive(name, RECEIVE_TIMEOUT);
           handleMessage(message);
           return;
        } catch (TimeOutException e) {}
} } }  \end{javap}
   \caption{{\Java} implementation of the {\RM} of the Two-Phase Commit protocol.}
   \label{fig:tp-java}
\end{figure}
 
An {\RM} is identified by a \inljava{name} and uses a
\inljava{network} (manager) to send and receive messages.  Once it
completes its task (represented by method \inljava{working}), it sends
a message to the {\TM} indicating that it is prepared to commit and
waits for a reply. The method \inljava{handleMessage} causes the
transaction to be committed or aborted, according to the decision
received from the {\TM}.  If no reply is received before the
\inljava{RECEIVE_TIMEOUT}, the {\RM} resends its prepared message to
the {\TM}.

\section{Instrumenting Distributed Programs}
\label{sec:instrument}

Our objective in this work is to check traces of program executions
against a \tlaplus specification of the algorithm the program is
expected to implement. In order to obtain such traces (in {\JSON}
format), we instrument implementations,
registering changes made to specification variables and logging
them to a file, together with a timestamp.
Section~\ref{sec:verify} describes the structure of \tlaplus trace
specifications used to process the traces generated in this way.

The main class of the {\Java} library we designed, \textsf{TLATracer},
essentially provides two methods: \inljava{notifyChange} for tracking
variable updates, and \inljava{log} used to
produce one log entry in the trace file that reflects all the variable
changes recorded with \inljava{notifyChange} since the last call to
\inljava{log}
(or since the start of the process if \inljava{log} was
never called before).\footnote{
A more detailed presentation of the library is given in the
appendix.
The library available at
\url{https://github.com/lbinria/trace_validation_tools/} offers more
convenience methods to track variable changes and log events.}

\begin{javap}
void notifyChange(String var, List<String> path,
                  String op, List<Object> args);  
long log(String eventName, Object[] args, long clockValue);
\end{javap}

Updates to variables are tracked using the method
\inljava{notifyChange}, which allows the programmer to specify 
an update to several fields (identified using the \inljava{path} argument)
of the \tlaplus variable \inljava{var} by applying the operation
\inljava{op} to the old field value and the arguments \inljava{args}.
Our library supports operations such as updating the variable by a new
value, adding or removing a value to or from a set or bag (multi-set) etc.
In its general form, the \inljava{log} method records the variable
changes as well as the name and optionally the parameters of the 
corresponding \tlaplus action.
The time when the log has been performed is used as a timestamp for
the corresponding entry. 
The library provides different types of clocks (in-memory, file-based or 
server-based) that can be used with the tracer, in which case the 
argument need not be indicated explicitly, but it should be
specified when the implementation uses a logical clock. The library
currently supports scalar logical clocks but vector clocks can easily be
added.

For example, in the \inljava{sendPrepared} method, an {\RM} sets its
state to \inljava{"prepared"} and sends a corresponding message to the
{\TM} (lines $2$ and $6$, respectively):\footnote{
We take some syntactic liberties, such as writing \inljava{\{...\}} instead of
\inljava{List.of(...)}.}
\begin{javap}
void sendPrepared()  {
  state = ResourceManagerState.PREPARED;
  tracer.notifyChange("rmState", {name}, "Update", {"prepared"});
  tracer.notifyChange("msgs", {}, "Add", {"type":"Prepared", "rm":name});
  tracer.log("RMPrepare", {name});
  networkManager.send(new Message(name, tmName, "Prepared", 0));
}
\end{javap}
The remaining lines are used for tracing purposes. Line~$3$ records the 
change of the entry corresponding to the {\RM} executing
\inljava{sendPrepared} ({\ie} \inljava{this.name})
of the \tlaplus variable \inlpluscal{rmState} to the new value 
\inlpluscal{"prepared"}. Similarly, line~$4$ indicates that a message of type 
\inlpluscal{"Prepared"} from the current {\RM} is added to the set \inlpluscal{msgs}.
Finally, in line~$5$ these changes are logged as corresponding to the
\tlaplus action \inlpluscal{RMPrepare} with the current {\RM} as parameter.
For an {\RM} named \inlpluscal{"rm-0"}, the above code produces
the following log entry in {\JSON} format:\footnote{The {\JSON} schema of a trace entry is given in the appendix.}

\smallskip
\noindent%
\{~\begin{tabular}[t]{@{}l}
  \str{clock}: $4$, \\
  \str{rmState}: [ 
    \{ \str{op}: \str{Update}, \str{path}: [\str{rm-0}], \str{args}: [\str{prepared}] \}
  ], \\
  \str{msgs}: [ \{ 
    \begin{tabular}[t]{@{}l}
      \str{op}: \str{Add}, \str{path}: [\,],\\
      \str{args}: [{\str{type}:\str{Prepared},\str{rm}:\str{rm-0}}]
    \} ],
    \end{tabular} \\
  \str{event}: \str{RMPrepare}, 
  \str{event\_args}: [\str{rm-0}]~\}
\end{tabular}

\smallskip

It should be noted that it is possible to trace updates to only a subset of 
specification variables. Similarly, indicating the name and the arguments of 
the \tlaplus action is optional.
For example, the size of the 
value of a variable might be prohibitively large, or the value might simply 
be unknown, for example because the implementation handles encrypted values.
In the above example, either or both calls to \inljava{notifyChange} could have 
been omitted. As discussed in Section~\ref{sec:verify}, the model checker 
will fill in suitable values for variables omitted from tracing. However, providing more detailed information can lead to more efficient validation and strengthen confidence in the results.

The library also provides \textsf{Python} scripts for merging the
trace files produced by different processes and for validating the
resulting trace using TLC.

\section{Checking Program Traces Against \tlaplus Specifications}
\label{sec:verify}

Having obtained a log from an execution of the distributed program, we must 
check if this log matches some behavior of the \tlaplus state machine
specification that the program is expected to implement. We define 
the problem more formally, explain how we realize trace validation,
report our experience with this approach,
and discuss some technical aspects of trace validation.

\subsection{The trace validation problem}
\label{sec:verify:strategy}

\begin{figure}[tb]
  \centering
    \begin{minipage}[t]{.33\linewidth}
    \centering

    Trace of implementation

    \bigskip

    \begin{tikzpicture}[>= stealth]
      \draw
      (0,-3.6) node [fill=green!80!black] (s1) {A}
      (0,-3.0) node [fill=red!60] (s2) {B}
      (0,-2.4) node [fill=orange!50] (s3) {C}
      (0,-1.8) node [fill=cyan!80!black, minimum size=5mm] (s4) {D}
      (0,-1.2) node [fill=yellow] (s5) {E}
      (0,-0.6) node [fill=magenta!50] (s6) {F}
      (0,0) node [fill=blue!30] (s7) {G}
      ;
      \draw[->] (s1) -- (s2);
      \draw[->] (s2) -- (s3);
      \draw[->] (s3) -- (s4);
      \draw[->] (s4) -- (s5);
      \draw[->] (s5) -- (s6);
      \draw[->] (s6) -- (s7);
    \end{tikzpicture}
  \end{minipage}
  \hfill
  \begin{minipage}[t]{.6\linewidth}
    \centering

    State space of \tlaplus specification

    \bigskip

    \begin{tikzpicture}[>= stealth, shorten >=1pt, node distance = 7mm, auto]
      \node[state,initial by arrow, initial text={}, fill=green!80!black, minimum size=6mm] (11) {A} ;
      \node[state, right=of 11, fill=red!60, minimum size=6mm] (12) {B};
      \node[state, right=of 12, fill=orange!50, minimum size=6mm] (13) {C};
      \node[state, right=of 13, fill=cyan!80!black, minimum size=6mm] (14) {D};
      \node[state, right=of 14, fill=yellow, minimum size=6mm] (15) {E};
      \node[state, above=of 11, fill=red!60, minimum size=6mm] (21) {B};
      \node[state, right=of 21, minimum size=6mm] (22) {};
      \node[state, right=of 22, fill=cyan!80!black, minimum size=6mm] (23) {D};
      \node[state, right=of 23, minimum size=6mm] (24) {};
      \node[state, right=of 24, fill=magenta!50, minimum size=6mm] (25) {F};
      \node[state, initial by arrow, initial text={}, above=of 21, fill=orange!50, minimum size=6mm] (31) {C};
      \node[state, right=of 31, minimum size=6mm] (32) {};
      \node[state, right=of 32, fill=yellow, minimum size=6mm] (33) {E};
      \node[state, right=of 33, fill=magenta!50, minimum size=6mm] (34) {F};
      \node[state, right=of 34, fill=blue!30, minimum size=6mm] (35) {G};

      \path[->]
        (11) edge (12) edge (21) edge (22)
        (12) edge (13) edge (22)
        (13) edge (14) edge (23)
        (14) edge (15) edge (24)
        (15) edge (24) edge (25)
        (21) edge (31)
        (22) edge (13) edge (21) edge (31) edge (32) edge (23)
        (23) edge (24) edge (33)
        (24) edge (25) edge (33)
        (25) edge (34) edge (35)
        (31) edge (32) 
        (32) edge (33)
        (33) edge (24)
        (33) edge (34)
        (34) edge (35)
        (31) edge[dashed] ++(0,1)
        (32) edge[dashed] ++(0,1)
        (33) edge[dashed] ++(0,1)
        (34) edge[dashed] ++(0,1)
        (35) edge[dashed] ++(0,1)
        (15) edge[dashed] ++(1,0)
        (25) edge[dashed] ++(1,0)
        (35) edge[dashed] ++(1,0)
     ;
    \end{tikzpicture}
  \end{minipage}
  \caption{Trace validation as a search for paths in the state space.}
  \label{fig:check-trace}
\end{figure}

Our problem can be stated as follows. Let $\mathcal{S}$ be the set of 
finite behaviors that satisfy the specification $Spec$, and let $\mathcal{T}$
be the set of finite behaviors represented by the trace, with arbitrary
values assigned to variables whose values are not recorded.
The trace is compatible with the specification if $\mathcal{S} \cap \mathcal{T} \neq \emptyset$. Note in particular that we do not check for refinement of the high-level 
specification by the trace, expressed as 
$\mathcal{T} \subseteq \mathcal{S}$: this condition would be too strong in the presence of
non-deterministically chosen values for variables not recorded in the trace.

Figure~\ref{fig:check-trace}
illustrates the idea. On the left-hand side, the chain of nodes
represents the trace obtained from the instrumented program. The graph on the
right-hand side represents the state space of the \tlaplus specification, and 
we must check if the trace can be matched to some path in the state space.
The first node of the trace must correspond to some initial state of the 
graph: in our example, we assume that this is the case for the state labeled~A in the 
lower left-hand corner. Then, we try to match at least one successor of an 
already matched state with the corresponding successor in the trace. There may
be several matching states, in particular due to incomplete information about 
variable values: in the example, we assume that two successors of state~A
match the second node of the trace. On the other hand, the state
labeled~C in the left-hand column of the state space does not have a successor
matching the node labeled~D in the trace. Overall, the trace of the example is 
compatible since there is at least one path in the state space that matches the trace.

The problem is actually a little more subtle: atomic steps 
of the implementation need not match precisely those of the specification
but may correspond to zero or several steps of the \tlaplus 
specification. We explain how we reduce the problem to one of constrained 
model checking and how we realize it using the TLC model checker.
In general, the programmer is in charge of tracking the correspondence between
the event names in the trace and the actions in the specification as
well as the correspondence between the names of the logged variables
and those from the specification. In the specific case presented in
the next section, these correspondences are one to one and could be
automatically generated.

\subsection{Realizing trace validation using the TLC model checker}
\label{sec:verify:tlc}

The set $\mathcal{S}$ of finite behaviors satisfying the original 
specification is defined by the \tlaplus formula $\mathit{Spec}$. In order to 
characterize the intersection $\mathcal{S} \cap \mathcal{T}$ of finite 
behaviors that also correspond to the 
trace obtained from the execution, we add constraints to $\mathit{Spec}$.
Concretely, our framework provides a module 
\inlpluscal{TraceSpec} that provides operators
for defining the constrained specification.

\medskip
\begin{pluscal}
------------------- MODULE TraceSpec -------------------
EXTENDS Naturals, Sequences, TLC, Json, IOUtils 
VARIABLE l
Trace == ndJsonDeserialize(IOEnv.TRACE_PATH)
IsEvent(e) == 
  /\ l \in 1 .. Len(Trace) /\ l' = l + 1 
  /\ "event" \in DOMAIN Trace[l] => Trace[l].event = e 
  /\ UpdateVariables(Trace[l])
\end{pluscal}

\medskip

The module declares a variable \inlpluscal{l} that will denote the number of the
current line of the trace. The definition of the operator \inlpluscal{Trace}
causes the {\JSON} representation of the trace to be internalized as a 
sequence of records whose fields correspond to the entries of the log file.
The operator \inlpluscal{IsEvent(e)} encapsulates processing 
the current line of the trace and 
generating the constraints imposed by it. It requires that \inlpluscal{l} 
is a valid index into the trace. The trace may explicitly indicate the event 
corresponding to the current transition by including an \inlpluscal{"event"}
field, in which case the operator checks for the expected value. (Any event 
parameters indicated by the entry are taken into account below.) The operator
increments the variable \inlpluscal{l} and computes new values
for the variables recorded in the current line of the trace, by evaluating the
operator \inlpluscal{UpdateVariables}:

\begin{pluscal}
  UpdateVariables(ll) ==
    /\ "rmState" \in DOMAIN ll => 
          rmState' = UpdateVariable(rmState, "rmState", ll)
    /\ ...  \* similar lines for variables tmState, tmPrepared, msgs
\end{pluscal}
  
That operator is defined as a conjunction that 
checks for each variable of the original specification if a corresponding entry 
exists in the current line of the trace and, if so, determines the new value of 
the variable from that entry.
The operator \inlpluscal{UpdateVariable} is predefined in our framework 
and computes the new value from the value of the first argument (i.e., the
unprimed variable) and the operator to be applied according to the trace. 
For example, the {\JSON} entry

\centerline{%
  \str{rmState}: [ 
    \{ \str{op}: \str{Update}, \str{path}: [\str{rm-0}], \str{args}: [\str{prepared}] \} 
  ]
}

\noindent
will give rise to the \tlaplus value 
\begin{pluscal}
[rmState EXCEPT !["rm-0"] = "prepared"]
\end{pluscal}
representing the function \inlpluscal{rmState} with the value at argument
\inlpluscal{rmState["rm-0"]} replaced by \inlpluscal{"prepared"}. A single 
update in the {\JSON} trace may correspond to changes to several parts of a 
complex value such as a function or a record. Predefined \tlaplus operators exist 
for the different operators that our framework currently supports, and this can be 
smoothly extended, both in the instrumentation library and at the \tlaplus level, 
should additional operators be desirable.

For every action of the original specification, we then 
construct a similar action of the trace specification by conjoining the predicate 
\inlpluscal{IsEvent}. 
For example, the action of the trace specification corresponding to the 
\inlpluscal{TMCommit} action of the specification of the two-phase commit 
protocol is defined as 

\begin{pluscal}
IsTMCommit == IsEvent("TMCommit") /\ TMCommit
\end{pluscal}

Because TLC evaluates formulas from left to 
right, the effect of these definitions is to first update state variables 
based on the information in the log and then evaluate the action predicate 
of the underlying specification. This evaluation checks that the 
predicate is satisfied while non-deterministically generating suitable 
values for any variables left open in the trace. 
For actions that take arguments, we define the constrained action such that 
any parameters provided in the trace indicate the instance of the original 
action that may occur:

\begin{pluscal}
IsTMRcvPrepared ==
  /\ IsEvent("TMRcvPrepared")
  /\ IF "event_args" \in DOMAIN Trace[l] /\ Len(Trace[l].event_args) >= 1
     THEN TMRcvPrepared(Trace[l].event_args[1])
     ELSE \E r \in RM : TMRcvPrepared(r)
\end{pluscal}

The overall next-state relation \inlpluscal{TraceNext} is defined as the 
disjunction of these actions. Writing the trace specification for a given 
algorithm specification is systematic and could be automated in most cases,
except when \inlpluscal{TraceNext} may include disjuncts corresponding to 
action composition as discussed in Sect.~\ref{sec:discrepancies}.

TLC evaluates the action \inlpluscal{TraceNext} from the current state in 
order to compute all possible successor states. Even if no such state can be 
found (as in the case of state~C in the leftmost column of Fig.~\ref{fig:check-trace}), 
there may still exist matching behaviors elsewhere in the state space. 
Therefore, we should not direct TLC to check for deadlocks of the constrained specification. 
It would also be inappropriate to verify the liveness property 
\inlpluscal{<>(l >= Len(Trace))},
which would not hold for incomplete prefixes of the constrained state space.
On the other hand, checking the invariant \inlpluscal{[](l < Len(Trace))} will 
cause TLC to output a behavior of the original specification that matches the trace 
if such a behavior exists. However, no useful information will be provided when the 
trace cannot be matched, i.e., when trace validation fails. 
Instead, we may observe that the
length of the longest path in the constrained state space 
corresponds to the diameter of the graph,\footnote{The
presence of the line counter \inlpluscal{l} excludes cycles in the 
constrained state space.}
which suggests defining the predicate 

\begin{pluscal}
TraceAccepted == TLCGet("stats").diameter - 1 = Len(Trace)
\end{pluscal}
as the postcondition to check for determining success of trace validation.
If the postcondition is violated, $\mathcal{S}$ does not contain any behavior 
whose prefix of appropriate length is in $\mathcal{T}$, i.e., matches the 
log of the execution. Because there is not a single behavior explaining this 
failure, TLC cannot generate a counter-example. However, it can generate a maximal
finite behavior in $\mathcal{S}$ that corresponds to the log but cannot be 
extended further. 
The \emph{hit-based breakpoint} 
feature of the \tlaplus debugger \cite{kuppeTLADebugger2023} can be used 
to halt state-space exploration when the diameter reported by TLC is attained,
and the user can then step back and forth in the state space in order to 
understand the discrepancy. 

Regardless of 
whether we use a property or a postcondition, trace validation might incorrectly 
accept a trace if the trace provides incomplete information. A particularly 
extreme case would be a trace that does not log any variable updates or 
action occurrences and only provides the length of a finite execution.

\subsection{Analyzing discrepancies}
\label{sec:discrepancies}

We have used trace validation for several case studies: besides the two-phase 
commit protocol presented here, we experimented with a distributed key-value 
store ensuring snapshot isolation~\cite{Fekete2009} whose specification was taken from the standard collection of \tlaplus
specifications~\cite{Lamport_TLA_Examples}, the distributed termination detection algorithm EWD~998 \cite{dijkstraShmuelSafraVersion1987,konnovSpecificationVerificationTLA2022a} from the 
same collection, two existing implementations of the Raft consensus algorithm~\cite{Ongaro2014},
and the Microsoft Confidential Consortium Framework \cite{howardConfidentialConsortiumFramework2023a}.
In all cases, trace validation quickly identified executions of the distributed 
implementations that could not be matched to the high-level specification. In the 
following, we identify several reasons for such discrepancies.

\para{Implementation shortcuts.}
The implementation of the Two-Phase Commit protocol could use a
counter to store the number (rather than the identities) of the {\RM}s
from which a \str{Prepared} message was received, and it could check if all
{\RM}s are prepared to carry out the transaction by simply comparing
the counter value to
the number of resource managers. Such an implementation will work correctly as
long as no messages are lost. However, if some {\RM} resends the \str{Prepared}
message due to a timeout, the {\TM} might count it twice and then commit 
prematurely. We found this kind of implementation error to be reliably detected 
using trace validation.

\para{Overly strict specification.}
The implementation~\cite{kuppeImplementingTLASpecification2023} of the token-based 
distributed termination detection algorithm EWD~998
allowed a node to send an ordinary (non-token) message to itself, which 
was ruled out in the specification. Again, trace validation quickly
discovered this mismatch, which can be resolved by adapting either the 
specification or the implementation.

\para{Mismatch of the grain of atomicity.}
Mainstream programming languages do not 
provide atomic transactions encapsulating several updates, so the choice of 
when to log an action corresponding to the \tlaplus specification requires 
consideration. Transitions that modify shared state such as sending or 
receiving network messages, acquiring or releasing global locks or committing 
state to stable storage are natural candidates. 
In most cases, we found 
it not too difficult to identify suitable points in the code for instrumentation,
but the choices are important for trace validation to be meaningful.

It frequently happens that the implementation
takes steps that are invisible at the level of the abstract state
space. In our Two-Phase Commit
running example, sent messages are (permanently) stored in a set, so resending a message 
due to a timeout conceptually corresponds to a stuttering step of the specification. 
However, when the implementation resends, say, a \inlpluscal{"Prepared"} message, it 
should not log an \inlpluscal{RMPrepare} step because such a step is not allowed to 
take place when the {\RM} is already in state \inlpluscal{"prepared"}. 
For this example, one can either 
choose to omit logging an action when a message is resent or only log the new 
(in fact unchanged) variable values but no action. The fact that \tlaplus 
specifications are insensitive to stuttering steps is helpful in such situations.

An implementation may also combine two or more separate steps of the
\tlaplus specification into a single transition. For example, in Raft 
implementations, nodes may update their term upon receiving an 
AppendEntries request for a higher term, whereas the two actions of updating 
the term and appending entries are 
distinguished in the Raft specification. In such cases, one may instrument 
the implementation such that it logs two transitions in succession.
Alternatively, one may add an explicit disjunct to the next-state relation 
of the \tlaplus specification used for trace validation, making use of the 
\tlaplus action composition operator $A \circ B$, support for which has 
recently been added to TLC. Doing so is a way of explicitly documenting
optimizations of the implementation with respect to the specification.

\subsection{Experience with trace validation}
\label{sec:experience}

Overall, we and the engineers we worked with found it quite straightforward
to instrument existing 
code, notably when the \tlaplus specification was used as a 
guideline for writing the implementation. The instrumentation library 
presented in Sect.~\ref{sec:instrument} with corresponding \tlaplus operators 
shown in Sect.~\ref{sec:verify:tlc} was found helpful but not strictly necessary:
in fact, it was not used in all of our case studies. Writing the trace specification is 
generally systematic and straightforward.

The most significant case study to which we applied trace validation occurred in 
the context of reverse engineering a formal specification for a Raft-inspired 
consensus algorithm implemented within the Confidential
Consortium Framework (CCF). The starting point was a specification written 
after analyzing the source code, which
was then corrected and amended through trace validation based on an existing test suite that exercises non-trivial system behavior. Model checking the resulting specification revealed serious violations of key safety properties. The counterexamples obtained in this way were translated into new tests, which confirmed previously unknown problems with the implementation. After addressing these issues at the specification level, corresponding updates to the implementation were made. To ensure ongoing consistency between the specification and its implementation, trace validation is now part of CCF's continuous integration pipeline. A detailed analysis of our experiences with formally verifying CCF, including trace validation, is discussed in a separate paper~\cite{howardSmartCasualVerification2024}.

We also leveraged trace validation when implementing the distributed termination detection algorithm EWD~998, starting from its preexisting \tlaplus specification. The algorithm is based on a token-passing scheme for detecting when global termination has occurred; its implementation consists of about 500 lines of code. Its specification includes an action for atomically passing the token from one node to its neighbor. The implementation is based on asynchronous message passing, and sending and receiving the token was logged as two separate transitions. Moreover, the implementation sends the token as soon as local termination occurs, whereas the specification has a separate action for termination detection. Trace validation pointed out both discrepancies, which are instances of differences in the grain of atomicity. Once these discrepancies were fixed by having the implementation log both termination detection and token sending, but not token receiving, the issue mentioned previously of a node possibly sending an ordinary message to itself was detected. Finally, it was found that the nodes continued to pass the token even after all nodes except for the initiator had terminated, which corresponded to an implementation error. Since then, thousands of traces have been successfully validated, and we are now confident that the implementation is indeed correct.

\subsection{Implementation aspects}
\label{sec:verify:implementation}

Our approach to instrumentation is flexible with regard to the detail of 
information recorded in the trace: only a subset of variables needs to 
be included in the trace, and names and parameters of corresponding \tlaplus
actions may or may not be given. Less information in the trace increases the 
potential degree of non-determinism in the trace specification and may
lead to a combinatorial explosion during model checking.
To some extent, this problem can be alleviated using different exploration
strategies. TLC's default breadth-first search (BFS) ensures
shortest-length counter-examples and is most informative for debugging. 
However, depth-first search (DFS) constrained by the length of 
the trace can be more efficient because checking can be stopped as soon as 
some behavior of the expected length has been found.

\begin{figure}[!t]
\centering
\begin{tabular}{|l@{~~}||@{~~}r@{~~}||@{~~}r@{~~}|@{~~}r@{~~}|@{~~}r@{~~}|@{~~}r@{~~}|@{~~}r@{~~}|}
\hline
   Instance & 
   \multicolumn{1}{c||}{length} &  
   \multicolumn{1}{c|}{VEA}  & 
   V         & 
   VpEA    & 
   EA      &  
   E \\
\hline
   \bTP{4}  &  17   &  19    & 211/35         &  19     & 48/22   & 246/58       \\
   \bTP{8}  &  33   &  35    & 8k/73          &  35     & 640/42   & 22k/695     \\
   \bTP{12} &  73   &  74    &  $\infty$/209  &  74     & 11k/86   & 2.5M/27k   \\
   \bTP{16} &  90   &  91    &  $\infty$/270  &  91     & 205k/107 & $\infty$/557k  \\
\hline
   \bKVS{4}{10}{20}   &  109   &  111   &  $\infty$/158  &   13k/149    &   111  &  $\infty$/35k  \\
   \bKVS{8}{10}{20}   &  229   &  231   &  $\infty$/317  &   18k/307    &   231  &  $\infty$/176k  \\
   \bKVS{12}{10}{20}  &  295   & 297    &  $\infty$/423  &   678k/411   &   297  &  $\infty$/300k  \\
\hline  
   \bKVS{4}{20}{40}   &  131   & 133   &  $\infty$/298   &   $\infty$/285   &   133  &  $\infty$/9.9M  \\
   \bKVS{8}{20}{40}   &  249   & 251   &  $\infty$/1164  &   $\infty$/1146  &   251  &  $\infty$  \\
   \bKVS{12}{20}{40}  &  308   & 310   &  $\infty$/552   &   $\infty$/538   &   310  &  $\infty$  \\
\hline
\end{tabular}
  \caption{Number of distinct states explored for valid traces
    of Two-Phase Commit (TP) and Key-Value Store (KV),
    for various degrees of precision (1hr timeout).}
  \label{fig:benchmarks}
\end{figure}

We report in Fig.~\ref{fig:benchmarks} the numbers of distinct states
explored with BFS/DFS by TLC for several valid traces that contain 
more or less information; a single figure indicates that BFS and DFS generate
the same number of states. We
consider traces for the Two-Phase Commit protocol for $4$, $8$, $12$,
and $16$ \RMs. 
For the Key-Value Store we consider $4$, $8$ or $12$ agents accessing 
a store with a maximum of either $10$ keys and $20$ values or $20$ keys and
$40$ values.
The column headings indicate the kind of information that was recorded 
in the traces: all variables and the events with their arguments (VEA),
just the variables (V), the variables and some events (VpEA),
only the events with their arguments (EA) or only event names (E).
For Two-Phase Commit, VpEA records only the events of the {\TM}, for 
Key-Value Store, only the start and end of transactions are logged.

As expected, tracing full information for
variables and the events requires exploring the least
number of states. Logging only the variables or only the 
event names
quickly leads to state-space explosion
that makes trace validation infeasible. However, recording well-chosen 
partial information is sufficient for limiting the state space.
For the Key-Value Store, logging events and their arguments is enough
because they uniquely determine the corresponding variable values.
For Two-Phase Commit, logging the events of the {\RM}s is unnecessary.

Besides an exponential growth of the number of states, too imprecise
logs may even lead to erroneous traces being accepted because the 
model checker may be able to infer suitable values that do not correspond 
to the actual ones. Nevertheless, such traces, which require little 
instrumentation, can still be useful at early stages of validation for 
finding issues. For instance, validating a trace containing
only the event names for $16$ {\RMs}
takes a considerable amount of
time but the bug concerning the implementation using counters mentioned in 
Sect.~\ref{sec:discrepancies} can still
be detected with such a trace in less than a minute.

\section{Related Work}
\label{sec:related}

The verification of execution traces against high-level properties or specifications has a long history in formal methods. Havelund \cite{havelundUsingRuntimeAnalysis2000} introduced runtime verification as a lightweight method for checking that a system's execution trace conforms to its high-level specification.  Runtime verification typically involves the generation of a monitor from the specification, which consumes the trace events to check conformance \cite{falconeTutorialRuntimeVerification2013}. Howard et al.~\cite{howardModelBasedTraceChecking2011} verified execution traces directly against the system's high-level specification.  Their work also demonstrated that it was feasible to use standard model checkers (ProB and Spin) to check execution traces against these specifications, a technique that we also employ. However, they did not consider distributed programs that require the use of (centralized or distributed) clocks for preserving causality, and they did not consider traces with incomplete information.

Tasiran et al.~\cite{tasiran2002} were the first to extract and validate traces obtained from a hardware simulator against a \tlaplus{} spec, demonstrating the practical applicability of trace validation.
The adoption of \tlaplus{} among distributed system practitioners, spurred by Newcombe et al.~\cite{Newcombe2015}, and the formalization of trace validation as a refinement check by Pressler~\cite{Pressler2020}, caused trace validation to be applied to real-world distributed systems.
For instance, Davis et al.~\cite{davisEXtremeModellingPractice2020} applied the technique to MongoDB, discovering a non-trivial implementation bug.
However, they faced challenges in consistently logging the implementation state, and aligning different grains of atomicity, which we attribute to them not leveraging \tlaplus's non-determinism to infer implementation state, and action composition to align atomicity. Niu et al.~\cite{niuVerifyingZookeeperBased2022} also validated traces of Zookeeper, ensuring that its implementation corresponds to its spec.
Similarly, Wang et al.~\cite{wangModelCheckingGuided2023} revealed several implementation bugs by replaying \tlaplus{} behaviors against instrumented implementations. Furthermore, the work by Wang et al. serves as an example of the challenges of aligning the grains of atomicity, illustrated by the authors asserting two bugs in a widely used and well-established specification \cite{Ongaro2014}. We contend that these are, in fact, common \tlaplus{} modeling patterns and can be handled with action composition.  Nevertheless, all efforts found non-trivial bugs in real-world systems by comparing implementation traces to high-level \tlaplus{} behaviors, a testament to the effectiveness of this approach.

To facilitate a closer alignment between high-level specifications and their actual implementations, Hackett et al.~\cite{hackettCompilingDistributedSystem2023} and Foo et al.~\cite{fooProtocolConformanceChoreographic2023} proposed extensions to PlusCal, an algorithmic language whose translator serves as a front-end to \tlaplus. These aim at adding sufficient detail to specifications for code generation and enabling the generation of runtime monitors, respectively. Despite these advancements, the requirement for implementation-level shims may impede widespread industry adoption.  Moreover, the projection from a global state machine, which is a common modeling pattern, to node-local state machines prevent the verification of global properties by monitors.  Yet, both approaches still translate specifications written in their PlusCal extensions into \tlaplus, allowing users to leverage all of the existing verification tools.

Notions of testing implementations against formal specifications, such as input-output conformance \cite{tretmans:test} are related to work on trace validation in that they also help establishing confidence in the correctness of implementations. A main difference is that in these approaches, implementations are considered black boxes that admit certain observations at the interfaces, whereas we assume having access to the implementation code. A similar remark holds for techniques of active learning of state machines~\cite{vaandrager:learning}. Although we only have anecdotal evidence, we observed that programmers find it quite easy to instrument their code, and that they will apply the necessary engineering judgment to overcome mismatches between the grains of atomicity of the specification and the implementation.

\section{Conclusion}
\label{sec:conclusion}

Formal verification techniques are known to be most effective for specifications 
written at high levels of abstraction where the size of state spaces 
(for model checking) and the complexity of invariants (for deductive
verification) are manageable. High-level specifications can serve as 
guidelines for programmers when implementing an algorithm, and in some 
cases it may even be possible to generate code from a sufficiently 
detailed specification. Model-based testing is a collection of 
techniques that rely on formal specifications for generating test cases, 
aiming at coverage guarantees or at exploring parts of the state space 
deemed interesting based on an analysis of the specification.

In this paper we described an approach for relating traces of the 
executions of a distributed program to the state machine described by a
high-level specification written in \tlaplus. Our purpose with this 
approach is to identify discrepancies that can be analyzed using the 
TLC model checker in order to determine if they are due to an error 
in the implementation, a restrictive specification, or an artefact due 
to a mismatch in the grains of atomicity. Although the technique does 
not provide formal correctness guarantees, mainly due to the necessary 
decisions on when to log an event corresponding to an atomic action, 
we have found it to be 
surprisingly effective for finding serious bugs in implementations 
that had previously been
validated using traditional quality assurance techniques.

We are certainly not the first to suggest that trace validation 
can be worthwhile for relating high-level specifications and 
programs. Original aspects, to the best of our knowledge, are our ability
to handle different grains of atomicity, and that we do not require
all specification variables or events to be traced in the log; instead, 
we use a model checker to reconstruct 
missing information. This leads to a tradeoff between the amount of 
detail included in the trace, the increase of the search space for 
model checking, and the reliability of the verdict. Our 
experiments suggest that it is enough to trace a suitably chosen
subset of variables and/or events. We implemented a library of 
{\Java} methods and \tlaplus operators to support collecting 
traces but their use is not strictly essential for applying our 
technique. In fact, the EWD~998 and the CCF case studies used ad-hoc 
code instrumentations for generating the logs.

We contribute industry-grade support for trace validation by implementing action composition, depth-first search, and debugging of trace validation in the TLC model checker and the \tlaplus debugger.
This support enabled not only the successful adoption of trace validation by the previously mentioned CCF project but also by the \textsf{etcd} project~\cite{etcd2024}, a widely-used distributed key-value store. Both projects verify that their implementations in C++ and Go, respectively, adhere to their \tlaplus specifications. We have exclusively used the explicit-state model checker TLC in our experiments. In principle, symbolic model checkers such as \tlaplus's Apalache tool \cite{10.1145/3360549} could also be used, but we suspect that the overhead of generating and evaluating constraints would be prohibitive in comparison to explicit-state model checking, in particular when the degree of non-determinism is low, as is the case when recording sufficiently informative traces.


At the moment, we exploit random variations in implementation parameters 
such as message delays or failures in order to generate meaningful traces 
of the program to be analyzed. In future work, we intend to use the 
\tlaplus specification in order to be able to steer executions towards 
``interesting'' parts of the state space. We also intend to study the 
feasibility of applying trace validation online during the execution 
of the program or even of using the technique as a run-time monitor 
to block unsafe transitions of the implementation.


\newpage
\appendix
\section{Specification of the  Two-Phase Commit protocol}

Fig.~\ref{fig:app:tp-tla} contains the \tlaplus specification of the
well-known Two-Phase Commit. In this version of the protocol, aborting
the transaction is left to the {\TM}.

\begin{figure}[!h]
\begin{pluscal}
CONSTANT RM
VARIABLES rmState, tmState, tmPrepared, msgs
vars == <<rmState, tmState, tmPrepared, msgs>>
TypeOK == rmState \in [RM -> {"working","prepared","committed","aborted"}]
Messages == [type : {"Prepared"}, rm : RM] \union [type : {"Commit","Abort"}]
TPInit == 
   /\ rmState = [r \in RM |-> "working"] 
   /\ tmState = "init" 
   /\ tmPrepared = {}
   /\ msgs = {}
RMPrepare(r) == 
   /\ UNCHANGED <<tmState, tmPrepared>>
   /\ rmState[r] = "working" 
   /\ rmState' = [rmState EXCEPT ![r] = "prepared"]
   /\ msgs' = msgs \union {[type |-> "Prepared", rm |-> r]}
RMRcvCommitMsg(r) == 
   /\ UNCHANGED <<tmState, tmPrepared, msgs>>
   /\ [type |-> "Commit"] \in msgs 
   /\ rmState' = [rmState EXCEPT ![r] = "committed"]
RMRcvAbortMsg(r) ==  
   /\ UNCHANGED <<tmState, tmPrepared, msgs>>
   /\ [type |-> "Abort"] \in msgs 
   /\ rmState' = [rmState EXCEPT ![r] = "aborted"]
TMRcvPrepared(r) == 
   /\ UNCHANGED <<rmState,tmState,msgs>> 
   /\ tmPrepared' = tmPrepared \union {r}
   /\ tmState = "init" 
   /\ [type |-> "Prepared", rm |-> r] \in msgs
TMCommit == 
   /\ UNCHANGED <<rmState, tmPrepared>>
   /\ tmState = "init"
   /\ tmPrepared = RM
   /\ tmState' = "done"
   /\ msgs' = msgs \union {[type |-> "Commit"]}
TMAbort == 
   /\ UNCHANGED <<rmState, tmPrepared>>
   /\ tmState = "init" /\ tmState' = "done" 
   /\ msgs' = msgs \union {[type |-> "Abort"]}
TPNext ==  
   \/ TMCommit \/ TMAbort
   \/ \E r \in RM : \/ RMPrepare(r) \/ TMRcvPrepared(r) 
                 \/ RMRcvCommitMsg(r) \/ RMRcvAbortMsg(r)
Spec == TPInit /\ [][TPNext]_vars     
Consistent ==
   \A r1,r2 \in RM: ~(rmState[r1] = "aborted" /\ rmState[r2] = "committed")
THEOREM Spec => [](TypeOK /\ Consistent)
\end{pluscal}
  \caption{\tlaplus Specification of Two-Phase Commit.}
  \label{fig:app:tp-tla}
\end{figure}

\section{Implementation of the Two-Phase Commit protocol}

A possible {\Java} implementation of the transaction manager and of
the resource managers is presented in Fig.~\ref{fig:app:tp-java}. Only
a simplified version of the main method is presented, the auxiliary methods are faithful
{\Java} translations of the actions in the {\tlaplus}
specification.

\begin{figure}[!h]
  \begin{javap}
public class TransactionManager {
  String name;
  Collection<String> resourceManagers, preparedRMs;
  NetworkManager network;
  public void run() throws IOException {
     while (true) {
        try {
           Message message = network.receive(name, RECEIVE_TIMEOUT);
           handleMessage(message);
        } catch (TimeOutException e) {}
        if (checkAllPrepared()) {
           commit();
           return;
        } else if (shouldAbort()) {
           abort();
           return;
} } }   }

public class ResourceManager {  
  String name, tmName;
  ResourceManagerState state;
  NetworkManager network;
  public void run() throws IOException {
     working();
     while (true) {
        sendPrepared();
        try {
           Message message = network.receive(name, RECEIVE_TIMEOUT);
           handleMessage(message);
           return;
        } catch (TimeOutException e) {}
} } }  \end{javap}
  \caption{{\Java} implementation of the managers of the Two-Phase Commit protocol.}
  \label{fig:app:tp-java}
\end{figure}

A {\TM} is identified by a \inljava{name} and uses a \inljava{network}
(manager) to send and receive messages. It stores the collection of
\inljava{resourceManagers} that it manages as well as the collection
of \inljava{preparedRMs} that have already indicated their
availability (empty at the beginning).
The {\TM} continuously reads messages and when the message corresponds
to a prepared {\RM}, the respective manager is added to
\inljava{preparedRMs} (in the method \inljava{handleMessage}). The
\inljava{receive} is blocking unless a timeout is reached.  When
all {\RMs} announced to be prepared, {\ie} \inljava{resourceManagers}
and \inljava{preparedRMs} contain the same elements (checked in method
\inljava{checkAllPrepared}), the {\TM} sends a message to each managed
resource manager (from \inljava{resourceManagers}) to inform them that
the transaction has been committed (method \inljava{commit}).
The {\TM} can decide to abort, for example because there are still
some {\RMs} who have not announced to be prepared before some
deadline, and in this case 
the {\TM} informs all the {\RMs} that
the transaction should be aborted (method \inljava{abort}).

\section{Detailed description of the instrumentation library}
\label{sec:app:instrumentApp}

The instrumentation library provides primitives that allow each
component of the implementation to log events reflecting its
behaviour.  The main class of the library, \textsf{TLATracer},
provides the methods for logging events and state variable updates:
\begin{javap}
static TLATracer getTracer(String tracePath, Clock clock);
static TLATracer getTracer(String tracePath);

void notifyChange(String var, List<String> path,
                  String operator, List<Object> args);  
VirtualField getVariableTracer(String var);

long log(String eventName, Object[] args, long clockValue);
long log(String eventName, Object[] args);
long log(String eventName)
long log();
\end{javap}

The method \inljava{getTracer} creates a tracer that logs events into
a file specified by the \inljava{tracePath} parameter. Each tracer
records the time of each event using a shared \inljava{clock},
ensuring that events are recorded in chronological order both locally
(within individual components) and globally (across all
components). While each component uses a unique \inljava{tracePath}
for its tracer, all tracers synchronize their timing using the same
type of clock. The library offers various types of clocks suitable for
different scenarios: an in-memory clock is used when components are
threads within the same process; a file-based clock is used for
processes on the same machine; and a server-based clock is appropriate
for distributed components. When a distributed clock is handled
explicitly by the components ({\eg} a Lamport or a vector clock) there
is no need to use a centralized clock; for this case the library
proposes the \inljava{getTracer} method with only one parameter.

Updates to variables are tracked by the method
\inljava{notifyChange}, which records operations that have been
applied to a given variable. The parameter \inljava{var}
refers to a variable from the {\tlaplus} specification but reflects
the operations executed at the implementation level and thus,
\inljava{notifyChange} implicitly links the variables from the
implementation to the ones in the specification. Our library supports
standard \inljava{operator}s such as updating the variable by a new
value, adding or removing a value to or from a set or bag (multi-set),
overriding the value of individual fields (identified using the
\inljava{path} argument) of functions or records, etc.  The path to
the field of the variable the operator applies on is specified as the
list of field names leading to it; for example,
\inljava{["address","city"]} to specify the city of residence of a
person having a field \inljava{address} which, in turn, has a field
\inljava{city}.  The list of arguments for the respective operator is
specified with \inljava{args}.

For example, the update of the control state of an {\RM} occurring in
the method \inljava{sendPrepared} could be recorded as follows (for
simplicity, in what follows, we write \inljava{\{...\}} instead of
\inljava{List.of(...)} and \inljava{new Object[]\{...\}}):

\begin{javap}
  state = ResourceManagerState.PREPARED; 
  tracer.notifyChange("rmState", {name}, "Update", {"prepared"});
\end{javap}

In the \tlaplus specification, the {\RM} states are modeled using the
\inlpluscal{rmState} function and thus, for an {\RM} named \emph{rm-0}
the above statement tracks the fact that \inlpluscal{rmState["rm-0"]}
is set to the new value \str{prepared}; the subsequent entry in the
final trace file would contain an element of the form:

\smallskip
\noindent%
\{~\begin{tabular}[t]{@{}l}
  \str{rmState}: [ 
    \{ \str{op}: \str{Update}, \str{path}: [\str{rm-0}], \str{args}: [\str{prepared}] \}
  ]
\end{tabular}

The \inljava{log} method is used to produce one log entry in the trace
file that reflects all the variable changes recorded with
\inljava{notifyChange} since the last call to \inljava{log} (or since
the start of the process if \inljava{log} was never called before). In
its general form, the \inljava{log} method records the variable
changes as well as the event name and its parameters, provided as
arguments of the \inljava{log} method. Variants of the \inljava{log}
method ignore the event or its parameters. The time when the log has
been performed is used as a timestamp for the corresponding entry;
this value is returned by the method.
The \inljava{clockValue} is only relevant if the (distributed) clock
is managed explicitly by the involved processes.  In our example,
since a centralized clock is used, the clock value doesn't need to be
specified explicitly and one of the other three versions of the
\inljava{log} method can be used.

For example, when a commit message is received by an {\RM}, the state
change is recorded using \inljava{notifyChanges} and the \inljava{log}
statement indicates that this corresponds in the specification to the
action \inlpluscal{RMRcvCommitMsg} for the corresponding {\RM}:
\begin{javap}
void handleMessage(Message message) {
  if (message.getContent().equals("Commit")) {
    state = ResourceManagerState.COMMITTED;
    tracer.notifyChange("rmState", {name}, "Update", {"committed"});
    tracer.log("RMRcvCommitMsg", {name});
  } ...
}  
\end{javap}
The trace entry obtained in the final trace file when executing the
above code for the {\RM} named \emph{rm-0} has the form

\smallskip
\noindent%
\{~\begin{tabular}[t]{@{}l}
  \str{clock}: $4$, \\
    \str{rmState}: [ \{ \str{op}: \str{Update},
      \str{path}: [\str{rm-0}],
      \str{args}: [\str{committed}] \} ], \\
    \str{event}:\str{RMRcvCommitMsg},\\
    \str{event\_args}:[\str{rm-0}]
    ~\}
\end{tabular}

The other variants of the method \inljava{log} ignore the arguments of
the event or even the event. For example, if in the method
\inljava{handleMessage} we used a simple \inljava{log} without
arguments then, the corresponding entry in the trace wouldn't contain
the lines \inljava{"event"} and \inljava{"event\_args"}.  As discussed
in Section~\ref{sec:verify:implementation}, more detailed information traced by
the implementation can lead a more efficient trace validation.

In general, the same variable is changed several times during the
execution and each change is recorded using the relatively verbose
\inljava{notifyChange} method. The library proposes a
\inljava{VirtualField} which can be used to trace the changes of a
variable, or of a field in a variable, in a more compact way; such a
variable tracer can be obtained using the method
\inljava{getVariableTracer} from \inljava{TLATracer}. To trace a
specific field of the variable, we can use the method
\inljava{getField} on the obtained \inljava{VirtualField}.

For example, we can obtain the tracers for the {\tlaplus} variables
\inlpluscal{msgs} and \inlpluscal{rmState}:
\begin{javap}
 VirtualField traceMessages = tracer.getVariableTracer("msgs");
 VirtualField traceStateRMs = tracer.getVariableTracer("rmState");
\end{javap}
and then the tracer for the field corresponding to \emph{rm-0} in \inlpluscal{rmState}:
\begin{javap}
 VirtualField traceState = traceStateRMs.getField("rm-0");
\end{javap}

A \inljava{VirtualField} can then be used to register a given operation on
the respective variable, potentially using some arguments:
\begin{javap}
  void apply(String op, Object... args){...}
\end{javap}
Several shortcut methods for the operators mentioned previously are
also provided: \inljava{update}, \inljava{add}, \inljava{remove},
\inljava{clear}, $\ldots$

For example, we can use the variable tracer defined above and replace
the line $4$ in the method \inljava{handleMessage} either with
\begin{javap}
   traceState.apply("Update", "committed");
\end{javap}
or with
\begin{javap}
   traceState.update("committed");
\end{javap}

As shown in Section~\ref{sec:instrument}, several variable changes can
be recorded successively and all these updates are combined into an
entry through a \inljava{log} call. We could have used the
\inljava{VirtualField}s in the \inljava{sendPrepared} method and use a
simpler \inljava{log} call (for simplicity, in what follows, we write
\inljava{k:v} instead of \inljava{Map.of(k,v)}):
\begin{javap}
void sendPrepared()  {
  state = ResourceManagerState.PREPARED;
  traceState.update("prepared");
  traceMessages.add({"type":"Prepared", "rm":name});
  tracer.log("RMPrepare");
  networkManager.send(new Message(name, tmName, "Prepared", 0));
}
\end{javap}

When executing the process for the {\RM} named \emph{rm-0}, the
tracing instructions lead to same line in the trace file as shown in
Section~\ref{sec:instrument} except for the \inljava{"event\_args"}
part:

\smallskip
\noindent%
\{~\begin{tabular}[t]{@{}l}
  \str{clock}: $4$, \\
  \str{rmState}: [ 
    \{ \str{op}: \str{Update}, \str{path}: [\str{rm-0}], \str{args}: [\str{prepared}] \}
  ], \\
  \str{msgs}: [ \{
    \begin{tabular}[t]{@{}l}
      \str{op}: \str{Add}, \str{path}: [], \\
      \str{args}: [{\str{type}:\str{Prepared},\str{rm}:\str{rm-0}}]
      \} ],
    \end{tabular} \\
  \str{event}: \str{RMPrepare}~\}
\end{tabular}

\section{JSON Schema for a trace entry}
\label{sec:app:schema}

An entry in the JSON file consists in the timestamp of the log and in
zero or several variable updates, each of which is identified by a key
having the name of the respective variable. Each variable update is an
array of objects, each of its elements consisting of the name of the
update operation, the array of fields indicating the path to reach the
targeted field the update is applied on, and the potentially empty
array of arguments for the update; an empty array of fields indicates
that the whole variable is updated.  The entry can also specify the
\str{event} concerned by the update and the arguments
(\str{event\_args}) of the corresponding action.

Fig.~\ref{fig:jsonSchema} presents the JSON Schema for an entry in the
trace file. The field \inljson{additionalProperties} stands for names of
variables logged in the entry.
\begin{figure}[!t]
\begin{json}
{
  "title": "TraceEntry",
  "description": "An entry in a trace",
  "type": "object",
  "properties": {
    "clock": {
      "description": "The timestamp of the event",
      "type": "integer",
      "minimum": 0
    },
    "additionalProperties": {
      "type": "array",
      "items": {
        "type": "object",
        "properties": {
          "op": { "type": "string" },
          "path": { "type": "array" },
          "args": { "type": "array" }
        },
        "required": [ "op", "path", "args"]
      },
      "minItems": 1
    },
    "event": {
      "description": "Name of the event",
      "type": "string"
    },
    "event_args": {
      "type": "array",
      "items": { "type": "string" } 
    }
    
  },
  "required": [ "clock"]
}
\end{json}

  \caption{JSON Schema for an entry in the trace file.}
  \label{fig:jsonSchema}
\end{figure}

\section{Debugging violations of $TraceAccepted$}
\label{sec:debugger}

\begin{figure}
 \centering
 \includegraphics[angle=90, width=\textwidth]{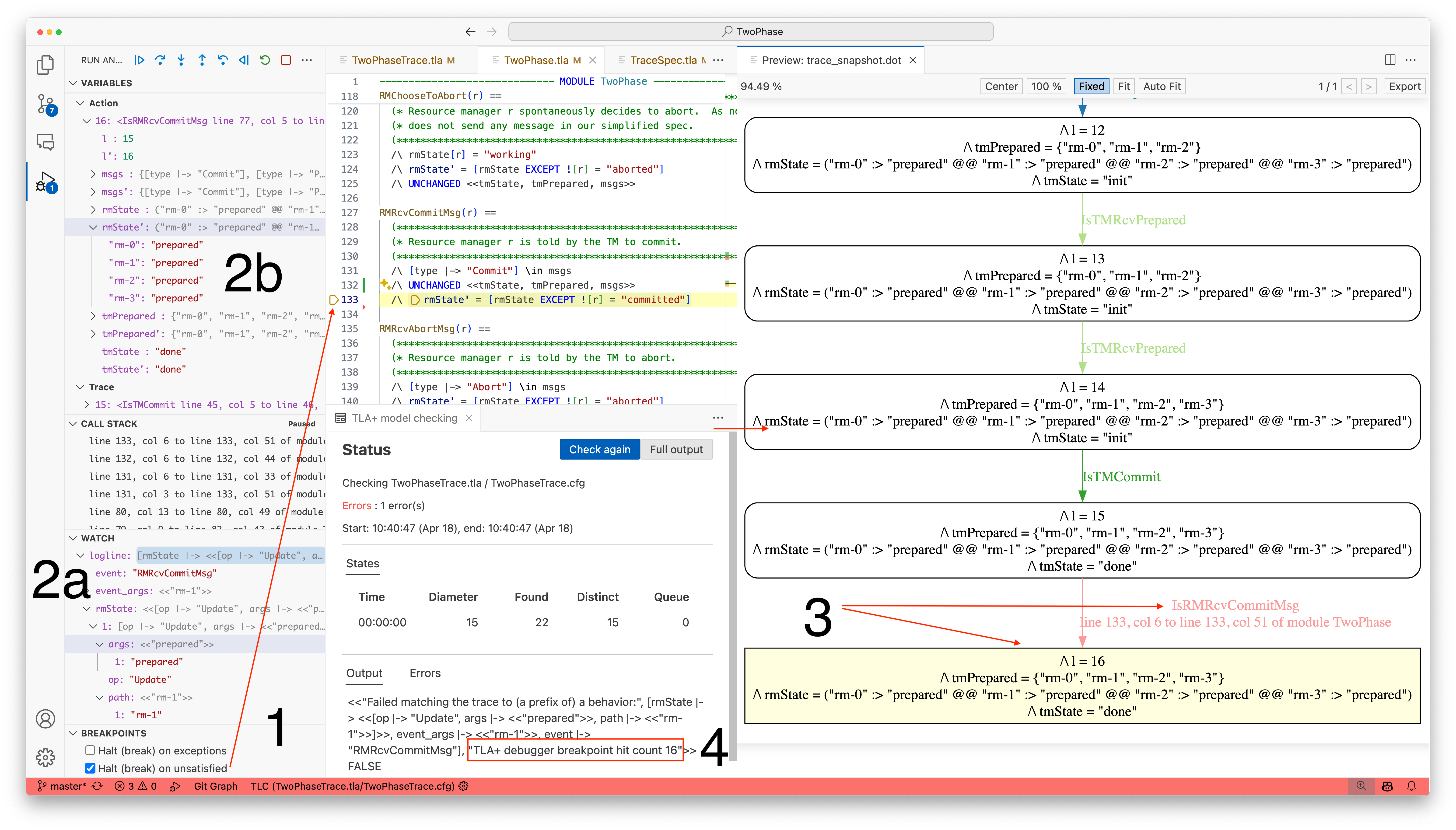}
 \caption{A screenshot of the \tlaplus debugger halted at the \textit{unsatisfied breakpoint}.}
 \label{fig:debuggerA}
\end{figure}

Section~\ref{sec:verify:tlc} introduced $TraceAccepted$, corrresponding to $\mathcal{S} \cap \mathcal{T} = \emptyset$, as the verification condition.  Contrary to ordinary model checking, a violation of the condition $TraceAccepted$ does not produce a counterexample. However, provided that $\mathcal{T} \neq \emptyset$, 
the behaviors within~$\mathcal{T}$ help explain why an implementation trace fails to conform to the specification. In instances of zero non-determinism in the constrained specification, where $\vert \mathcal{T} \vert = 1$, it is generally advisable to examine the final state of the behavior and the corresponding line in the trace to identify the source of the mismatch. For more complex specifications involving multiple variables and actions, the \textit{hit-based breakpoint} feature of the \tlaplus debugger can be used to halt state-space exploration when the diameter reported by TLC matches the value of a reported violation. Once halted, the debugger allows the user to step back and forth through the evaluation of action formulas to pinpoint the discrepancy. Additionally, the \tlaplus debugger displays the values of variables at both the current and successor states, facilitating a comparison with the trace values. With $\vert \mathcal{T} \vert > 1$, we provide a new \textit{unsatisfied breakpoint} that activates for each state in $\mathcal{T}$ that is found to be unreachable. Furthermore, $\mathcal{T}$ can be visualized as a graph that not only includes all unreachable states but also references the subformula responsible for the state being unreachable.

In figure~\ref{fig:debuggerA}, the \tlaplus debugger reached the \textit{unsatisfied breakpoint} that was triggered when the conjunct on line 133 evaluated to false ($1$).  The conjunct evaluated to false because the ({\JSON}) trace indicated a \inlpluscal{"RMRcvCommitMsg"} action ($2a$) that changed the state of resource manager \inlpluscal{"rm-1"} to \inlpluscal{"prepared"} ($2b$).  However, the Two-Phase specification on line 133 defined the \inlpluscal{"RMRcvCommitMsg"} action to change the state of the resource manager \inlpluscal{"r"} to \inlpluscal{"committed"}.  The view on the right shows the partial state graph up to the unreachable state, and the indicator that the formula on line 133 evaluated to false ($3$).  The "Status" view at the bottom displays the previous TLC run, which reported the violation of $TraceAccepted$ at a diameter of 16 ($4$).

\end{document}